\documentclass[aps,prl,twocolumn,unsortedaddress]{revtex4}%
\usepackage{amsmath}
\usepackage{graphicx}%
\usepackage{amsfonts}%
\usepackage{amssymb}
\begin{document}

\title{Dimensional cross-over and charge order in half-doped Manganites and Cobaltites}
\author{Oron Zachar$^{1,2}$ and Igor Zaliznyak$^{1}$}
\affiliation{$^{1}$Brookhaven National Laboratory, Upton, New York 11973-5000 USA. $\ $}
\affiliation{$^{2}$UCLA Dep. of Physics, Los Angeles, CA 90095, USA.}

\begin{abstract}
We propose a generic model for understanding the effect of quenched disorder on
charge ordering in half-doped Manganese- and Cobalt-Oxides with different
crystal structures. Current experimental observations (Table
\ref{LaSrCoO-ChargeOrder}) are explained in the light of the global phase
diagram of the model (Fig. \ref{RFI-PhaseDiagram}).

\end{abstract}


\maketitle

In the past decade much theoretical work was devoted to elucidating the nature
and origin of charge ordering (CO) in the doped transition metal oxides,
particularly in manganites, which are of great experimental and practical
interest
\cite{Millis-ManganiteOrder,Daggoto01-ManganiteTheoryReview,01-ManganiteExperimentalReview}.
Much attention was focused on studying the interplay of charge correlations with
orbital and spin degrees of freedom which are often important for the CO
\cite{TokuraNagaosa2000}. Surprisingly, however, no coherent approach to
understanding the effect of quenched disorder, such as introduced by a random
distribution of dopant ions, on charge order (distinguished from orbital order)
has been proposed so far.

Dopant disorder is a source of an unavoidable random electrostatic potential
which couples linearly to charge density fluctuations
\cite{01-ManganiteExperimentalReview}. In a strongly correlated electron system
close to criticality such perturbation is of crucial importance. We propose that
essential account for this random potential can be formulated in terms of a
generic model (\ref{RFIM}), mapping the CO interaction on an effective Ising
problem. We construct the phase diagram of this model, Fig.
\ref{RFI-PhaseDiagram}, and find that it adequately describes the essential
features of the CO structure in half-doped manganese and cobalt oxides. In
particular, a dimensional cross-over in our model gives a natural explanation to
the puzzling experimental observation that CO in the layered perovskites is
short-range \cite{Sternlieb1996,Zaliznyak2000,Wakabayashi2001_LaSr1.5MnO}, while
it is long-range in the pseudo-cubic materials
\cite{Radaelli1997_LaCaMnO3,Zimmermann99_PrCa0.5MnO,Nakamura1999_NdSrMnO3}.

\begin{figure}[pb]
\vspace{-0.25in}
\begin{center}
\includegraphics[width=3.4in]
{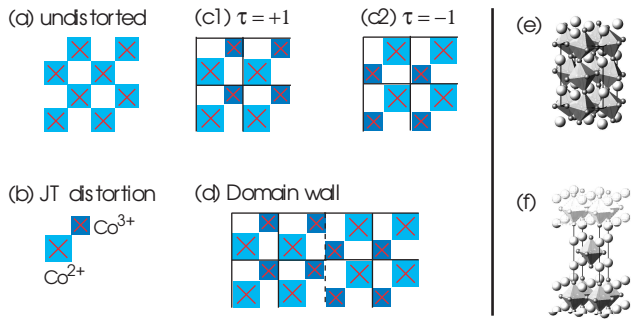}%
\caption{Charge order and JT distortion at half doping: $ab$ plane is tiled with
cells $\left\{\mathbf{r}\right\}$ containing two cobalt (manganese) ions,
(a)-(d). Crystal structure of pseudo-cubic perovskite material \mbox{${\rm
Pr_{0.5}Sr_{0.5}MnO_3}$} (e) and layered \mbox{${\rm La_{1.5}Sr_{0.5}CoO_4}$}
(f).}
\label{LaSrCoO-ChargeOrder}%
\end{center}
\vspace{-0.3in}
\end{figure}

At half-doping, the ground state of Co oxides and of many isostructural Mn
oxides manifests a checkerboard-type planar charge modulation accompanied with a
Jahn-Teller (JT) distortion of the MO$_{6}$ (M = Mn, Co, \textit{etc.})
octahedra shown in Fig. \ref{LaSrCoO-ChargeOrder},
\cite{Sternlieb1996,Zaliznyak2000,Wakabayashi2001_LaSr1.5MnO,
Radaelli1997_LaCaMnO3,Zimmermann99_PrCa0.5MnO,Nakamura1999_NdSrMnO3,comment1}.
The electrons are highly localized in a charge-ordered state with alternating
Co$^{2+}$\ and Co$^{3+}$\ valence (Mn$^{4+}$ and Mn$^{3+}$ in manganites), and
the in-plane O$^{2-}$ ions move towards the higher-valence M ions. To be
explicit, we present the model construction in terms of a
La$_{1.5}$Sr$_{0.5}$CoO$_{4}$ \cite{Zaliznyak2000} which has no superimposed
further symmetry breaking due to orbital order.

As a general framework, we argue that hopping dynamics, orbital, magnetic, local
Coulomb and Jahn-Teller interactions can all be integrated out into an effective
charge-ordering interaction which then competes with the random charge potential
introduced by the dopant ions.
A perfectly ordered state has a two-fold degeneracy associated with the choice
of sublattice occupied by the higher/lower valence ions. We define a local Ising
variable, $\tau\left( \text{\textbf{r}}\right)  =\pm1$, to denote the two
alternative ionic configurations, Fig. \ref{LaSrCoO-ChargeOrder}(c). A
single-domain CO state corresponds to a ferromagnetic order of $\tau\left(
\text{\textbf{r}}\right) $. Topological disorder is introduced by domain walls
as shown in Fig. \ref{LaSrCoO-ChargeOrder}(d). In principle, the model can be
enriched by introducing the `vacancies', $\tau\left( \text{\textbf{r}}\right)
=0$, which represent cells with even number of electrons, but we shall exclude
such possibilities in the analysis and discussion of this paper.

Because of the Jahn-Teller distortion, there is an elastic strain energy cost
for nearest neighbor cells with opposite $\tau$ values, Fig.
\ref{LaSrCoO-ChargeOrder}(d). This strain energy can be described by an
effective short-range ferromagnetic interaction $J \left( \mathbf{r-r}^{\prime}
\right)  \tau\left(  \text{\textbf{r}}\right)  \tau\left(
\mathbf{r}^{\prime}\right)  $ \cite{JahnTeller-Review75}. Without a loss of
generality we may approximate $J\left( \mathbf{r-r}^{\prime}\right)$ by a
nearest-neighbor coupling. The potential introduced by dopant ions randomly
favors one or the other configuration of a two-ion cell, and therefore
effectively acts as a random field $h\left(  \text{\textbf{r}}\right) \tau\left(
\text{\textbf{r}}\right)  $. Importantly, depending on the crystal structure the
effective coupling may be anisotropic, $J_{c}\neq J_{ab}$. Hence, for the charge
order, the effect of elastic crystal strain and random dopant distribution is
captured by an effective \emph{anisotropic} random-field-Ising-model (RFIM),
\begin{equation}
\label{RFIM}%
H=-J_{ab}\sum_{\left\langle \mathbf{rr}^{\prime}\right\rangle _{ab}}%
\tau_{\mathbf{r}}\tau_{\mathbf{r}^{\prime}}-J_{c}\sum_{\left\langle
\mathbf{rr}^{\prime}\right\rangle _{c}}\tau_{\mathbf{r}}\tau_{\mathbf{r}%
^{\prime}}+\sum_{\mathbf{r}}h_{\mathbf{r}}\tau_{\mathbf{r}}\;,
\end{equation}
where $\left\langle \mathbf{rr}^{\prime}\right\rangle _{ab}$ and $\left\langle
\mathbf{rr}^{\prime}\right\rangle _{c}$ denote nearest-neighbor cells in the
$ab$-plane and along the $c$-axis respectively, see Fig.
\ref{LaSrCoO-ChargeOrder} (e), (f). To our knowledge, all previous analytical
and numerical studies focused on the isotropic $3D$ RFIM. Here we make the first
attempt aimed at a qualitative and quantitative understanding of the anisotropic
RFIM (\ref{RFIM}).

\begin{figure}[pt]
\begin{center}
\includegraphics[height=2.0289in, width=2.5374in]
{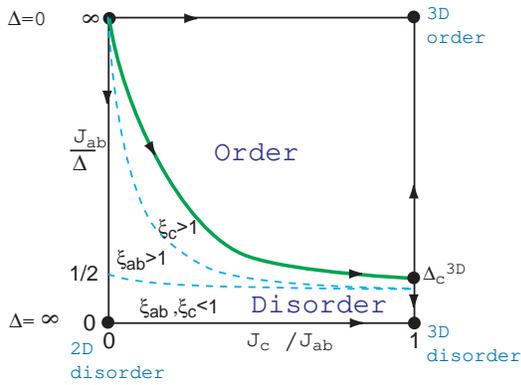}%
\caption{Charge order phase diagram from the anisotropic random field Ising
model (\ref{RFIM}). $J_{c}/J_{ab}$ and $J_{ab}/\Delta$ parameterize the
anisotropy of the effective coupling and the relative strength of the disorder
potential from dopant ions, respectively.}
\label{RFI-PhaseDiagram}%
\end{center}
\vspace{-0.3in}
\end{figure}

Random field alters both lower critical dimension \cite{Imry-Ma1975_Disorder}
and critical exponents \cite{RFI-NewExponents} of an Ising system. In
particular, for weak random field, $\Delta/J\ll1$, the isotropic $3D$ RFIM
orders at a finite temperature $T_{c}>0$, while the $2D$ RFIM has no long range
order (LRO) even at $T=0$ ($J$ and $\Delta=\left\{ \text{rms of
}h_{\mathbf{r}}\right\}$ parameterize the Ising coupling and disorder potential,
respectively). This prompts an interesting question: what is the fate of a $3D$
anisotropic random field Ising model (ARFIM) of (\ref{RFIM})? The simplest way
of connecting the renormalization group (RG) flows leads us to the phase diagram
shown in Fig. \ref{RFI-PhaseDiagram}.

To connect our model analysis with the experiment, we summarize the charge order
correlation lengths in the $ab$-plane, $\xi_{ab}$, and along the $c$-axis,
$\xi_{c}$, measured in several representative perovskite oxides in Table
\ref{LaSrCoO-ChargeOrder}. Correlations in the pseudo-cubic systems I-III (Fig.
\ref{LaSrCoO-ChargeOrder} (e)) are resolution-limited (typically this means
$\xi\gtrsim2000$ \AA\ \cite{Radaelli1997_LaCaMnO3,Zimmermann99_PrCa0.5MnO}), and
CO is apparently long-range. On the other hand, in the layered systems IV-V
(Fig. \ref{LaSrCoO-ChargeOrder} (f)) the CO is finite-range, with highly
anisotropic $3D$ correlations $\xi_{ab} \gg\xi_{c}$, which implies $J_{ab} \gg
J_{c}$. Large in-plane correlation length $\xi_{ab}\gg1$ indicates weak planar
disorder, $\Delta\ll J_{ab}$. Therefore, these materials probe the most
interesting and complicated region of the phase diagram.

\begin{table}[b]
\vspace{-0.25in}
\caption{Correlation lengths in terms of number of unit cells
in each direction (i.e. $\xi_{c}$ in units of $c\approx12.5$ \AA, and $\xi_{ab}$
in units of $a\sqrt{2}\approx5.44$ \AA, as implied by Fig.
\ref{LaSrCoO-ChargeOrder}).}%
\label{table1}
\vspace{0.05in}
\begin{tabular}
[c]{lcccc}\hline
\# & Material & Ref. & $\xi_{ab}$ & $\xi_{c}$\\\hline\hline
I. & La$_{0.5}$Ca$_{0.5}$MnO$_{3}$ & \cite{Radaelli1997_LaCaMnO3} & $\infty$ &
$\infty$\\
II. & Pr$_{0.5}$Ca$_{0.5}$MnO$_{3}$ & \cite{Zimmermann99_PrCa0.5MnO} &
$\infty$ & $\infty$\\
III. & Nd$_{0.5}$Sr$_{0.5}$MnO$_{3}$ & \cite{Nakamura1999_NdSrMnO3} & $\infty$
& $\infty$\\
IV. & La$_{0.5}$Sr$_{1.5}$MnO$_{4}$ & \cite{Sternlieb1996} & $\approx35$ &
$\approx3.3$\\
V. & La$_{1.5}$Sr$_{0.5}$CoO$_{4}$ & \cite{Zaliznyak2000} & $\approx4$ &
$\lesssim1$\\\hline
\end{tabular}
\end{table}

Because the isotropic $3D$ RFIM is long-range ordered for $\Delta <
\Delta_c^{3D} \approx 2.27 J$ \cite{Fisher01-RFIM3D-numerical}, it is clear that
for $\Delta < J_{c} < J_{ab}$ the system is in the regime of weak $3D$ disorder
and has a $3D$ LRO. Thus, we examine an intriguing case,
\begin{equation}
\label{J_c << d << J_ab}%
J_{c}\ll\Delta<J_{ab}\;,
\end{equation}
where the $c$-axis coupling is in the regime of \emph{strong} disorder, while
the $ab$-planes are in the \emph{weak} disorder limit.

For $J_{c}=0$ the system decouples into independent $2D$ RFIM planes, and the
ground state is disordered. This form of disordered ground state is
perturbatively stable. For an infinitesimal inter-plane coupling
$J_{c}/\Delta\ll1$, the $ab$-planes remain uncorrelated on all length scales
($c$-axis correlation length $\xi_{c}<1$), and each plane may still be treated
as an effectively independent $2D$ RFIM. Therefore, a critical line of phase
transitions exists in the $(J_{ab}/\Delta,J_{c}/J_{ab})$ phase diagram,
separating the disordered and $3D$-LRO phases (solid line in Fig. 2). With
increasing $J_{c}/J_{ab}$ on the disordered side we enter a critical regime,
where the ground state is still disordered but with significant and highly
anisotropic correlations in all spacial directions. In the isotropic $2D$ and
$3D$ RFIM \cite{Villain-RFIM_review} the ground state correlation length scales
with $\kappa=\Delta/J$ as $\xi ^{2D}\approx\exp\left(
\frac{1}{\sigma}\kappa^{-2}\right)  $, and $\xi ^{3D}\approx\left(
\kappa-\kappa_{c}\right)  ^{-\nu}$ (for $\kappa\geq \kappa_{c}$), respectively.
What are the correlation lengths in a highly anisotropic $3D$ RFIM (\ref{RFIM}),
with $J_{c}\ll\Delta<J_{ab}$?

The phase diagram, Fig. \ref{RFI-PhaseDiagram}, instructs that upon a RG
transformation model (\ref{RFIM}) scales either to a $3D$ LRO fixed point, or
towards a $3D$ strong disorder fixed point. Currently, there is no generally
satisfying analytical derivation of the RG equations for the RFIM for $d<4$.
However, because $d=2$ is the lower critical dimension of the RFIM, 3D random
field critical fixed point should be accessible perturbatively in
$2+\varepsilon$ expansion, \cite{Cardy-RGbook}. Therefore, we derive
perturbative RG equations for the anisotropic $3D$ RFIM (\ref{RFIM}) starting
from a semi-phenomenological scaling "ansatz" for the $2D$ RFIM
\cite{Villain-RFIM_review,Cardy-RGbook}. These equations, which naturally
account for the RG flow of the coupling anisotropy, allow us to derive and
analyze all qualitative features of the phase diagram of Fig.
\ref{RFI-PhaseDiagram}.

For the isotropic $d$-dimensional RFIM a simple scaling analysis of Imry-Ma,
$\left\{ J(l) \sim Jl^{d-1};\; \Delta\left( l \right) \sim \sqrt{l^{d}} \Delta
\right\}$, \cite{Imry-Ma1975_Disorder,Cardy-RGbook}, predicts the relative
disorder $\kappa=\frac{\Delta}{J}$ to scale as
\begin{equation}
\label{kappa}%
\kappa\left(  l\right)  =\frac{\Delta\left(  l\right)  }{J\left( l\right)
}\approx\frac{\sqrt{l^{d}}}{l^{d-1}}\kappa \;.
\end{equation}
The marginality of $\kappa\left(  l\right)  $ in 2D is associated with $d=2$
being the lower critical dimension. It is broken by subdominant contributions
from the interface roughening at the domain boundaries which lead to a
disordered ground state \cite{Cardy-RGbook,Villain-RFIM_review}. For weak
disorder, $\kappa\ll1$, the correlation length is $\xi=A\exp\left\{
+\frac{1}{\sigma}\kappa^{-2}\right\}$, with $\sigma\approx1$ (clearly, this
expression fails for strong disorder, where $\xi\leq1$). By demanding that
$\kappa\left( l \right)$ scales consistently with the correlation length,
$\xi\left(  l\right) = \xi/l$, and with the expression $\xi\left( l\right)
=A\exp\left\{ +\frac{1}{\sigma}\kappa^{-2}\left( l\right) \right\}$, we obtain a
scaling equation
\begin{equation}
\label{2D scaling Ansatz}%
\kappa^{-2}\left(  l\right)  = \kappa^{-2} \left[1 - \kappa^{2}\sigma\ln\left(
l\right)\right] \;,
\end{equation}
i.e. scaling of the relative disorder in 2D is governed by the logarithmic
corrections to Eq. (\ref{kappa}). The disordered ground state is manifested by
the flow of $\kappa^{-1}\left( l\right) =\frac{J\left(  l\right) }{\Delta\left(
l\right) }\rightarrow0$. In the context of the model (\ref{RFIM}), equation
(\ref{2D scaling Ansatz}) describes the RG flow for $J_{c}=0$.

To proceed with deriving the qualitative features of the RG flows shown in Fig.
\ref{RFI-PhaseDiagram} for $0<J_{c}\leq J_{ab}$ we note that the anisotropy of
interactions is, in fact, an irrelevant perturbation at the 3D critical point.
This is reflected by the fact that all stable fixed points are located on the
isotropic line in the phase diagram, at $J_{c}/J_{ab}=1$. At the mean-field
level (and without random fields) the anisotropy of the interactions can be
removed by appropriate anisotropic re-scaling of the coordinates
\cite{Cardy-RGbook}. Here we implement this idea by devising an
\emph{anisotropic} real-space RG transformation, effectively coarse-graining to
anisotropic blocks $l_{ab}\times l_{ab}\times l_{c}$. We derive the RG equations
along the lines of $2+\varepsilon$ expansion, and demonstrate explicitly the
flow towards the isotropic fixed points.

We parametrize the anisotropy as $\alpha = \left( \frac{J_{c}} {J_{ab}}
\right)^{x}$, $x > 0$, and define the RG transformation through an infinitesimal
anisotropic re-scaling, $\{ d(\ln l_{ab}) = d(\ln l) \;;\; d(\ln l_{c}) =
\alpha(l) d(\ln l) \}$. This determines the shape of the coarse-grained blocks
to be gradually varying, flowing from the anisotropic towards isotropic. We use
the Imry-Ma approximation \cite{Cardy-RGbook} to derive the lowest order
contributions to the flow equations for the model parameters, and supplement it
by the appropriate higher order correction to reproduce the scaling equation
(\ref{2D scaling Ansatz}) for the 2D RFIM case, $\alpha=0$. The resulting
perturbative RG equations are
\begin{align}
\frac{\mathrm{d}\left( \kappa^{2}\right)}{\mathrm{d}\ln\left( l\right) } & =
-\alpha\kappa^{2}+\sigma\kappa^{4}+\mathcal{O}\left(  \alpha^{2}\right)
\label{RG1}\\
\frac{\mathrm{d}\alpha}{\mathrm{d}\ln\left( l\right) } & = x \left( 1 -
\alpha\right) \alpha
\label{RG2}%
\end{align}
We obtain an unstable 2D fixed point at $\{\alpha^{\ast} = 0, \; \kappa^{\ast 2}
= 0 \}$, and stable isotropic 3D fixed points at $\alpha^{\ast}=1$. The $3D$
critical point is at $\{\alpha^{\ast} = 1, \; \kappa^{\ast 2} = \kappa_{c}^2 =
{\alpha^{\ast}/\sigma} \}$. With $\kappa_{c} \approx 2.27$ from numerical
simulations \cite{Fisher01-RFIM3D-numerical}, the extrapolation of our
perturbative equations to $\alpha=1$ would imply $\sigma\approx0.2$, which is
too small. This indicates that while qualitatively correct, and valid in the
perturbative region of parameters, RG equations (\ref{RG1}),(\ref{RG2}) are
inadequate for quantitative estimates.

It is straightforward to integrate Eq. (\ref{RG2}) and obtain $\alpha(l)^{-1} =
1 + (\alpha(1)^{-1} - 1) l^{-x}$, the scaling flow of the anisotropy. In
principle, the phase transition line can be determined by a further numerical
integration. The important asymptotes can be derived analytically,
\begin{align}
&\alpha (\kappa^2) \approx C_0\; e^{-\frac{x}{\sigma\kappa^{2}}}\;, & \alpha \ll
1\;,
\label{PD1}\\
&\alpha (\kappa^2) \approx 1 + C_1\; (\kappa^{2} - \kappa_c^{2})\;, & 1 - \alpha
\ll 1\;,
\label{PD2}%
\end{align}
and show that neglecting $\mathcal{O}\left(  \alpha^{2}\right)$ term in Eq.
(\ref{RG1}) is indeed justified for $\alpha \ll 1$, but not for $\alpha \approx
1$.

\begin{figure}[ptb]
\begin{center}
\includegraphics[height=0.6685in,width=2.6039in ]
{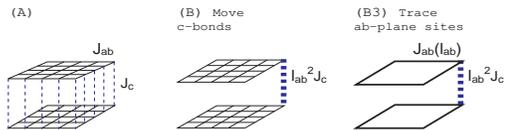}%
\caption{Block transformation procedure}%
\end{center}
\vspace{-0.3in}
\end{figure}

To obtain quantitative estimates of the anisotropic correlation lengths as a
function of $J_{c}$, $J_{ab}$ and $\Delta$, we use an alternative approach. In
the absence of the exact RG equations for the 3D ARFIM, we resort to a
real-space rescaling procedure where we employ the Migdal-Kadanoff bond-moving
technique \cite{Kadanoff76-BondMoving} combined with the scaling ansatz (\ref{2D
scaling Ansatz}). We adopt conventional approximations in ignoring the
generation of longer range interactions \cite{Villain-RFIM_review} and using the
``majority rule'' \cite{Kadanoff76-BondMoving} for block spin variable. Hence,
we follow only the transformation of effective nearest neighbor block
interactions $J_{ab}(l)$, $J_{c}(l)$ and effective disorder parameter
$\Delta(l)$.

Our strategy is to map the anisotropic RFIM onto a ''solvable'' isotropic $3D$
RFIM, whose known properties enable us to derive the estimates for
$\xi_{ab}\left( J_{ab},J_{c},\Delta\right)  $ and $\xi_{c}\left(
J_{ab},J_{c},\Delta\right)$. Because the correlation length scales as $\xi\left(
l\right) =\xi/l$, a \emph{bare anisotropic} $3D$ model with finite
$\xi_{ab}>\xi_{c}\geq1$ is mapped onto an \emph{effective isotropic} $3D$ model
with $\xi_{ab}=\xi_{c}$ when the coarse-grained block size $l_{ab}\times
l_{ab}\times l_{c}$ is chosen with $l_{ab}=\xi_{ab}/\xi _{c}$ and $l_{c}=1$. In
other words, the cross-over from quasi-$2D$ to isotropic $3D$ scaling is
intuitively implemented in two steps. First, a $2D$ transformation to blocks of
size $l_{ab}\times l_{ab}\times1$, and then a usual isotropic $3D$ scaling (see
Fig. 3).

Performing the Migdal-Kadanoff transformation to $l_{ab}\times l_{ab}\times1$
blocks we first move the $J_{c}$ bonds \cite{Kadanoff76-BondMoving} and thus
create an effective block interaction $J_{c}\left(  l_{ab}\right)
=J_{c}l_{ab}^{2}$, Fig. 3B. Now that the c-axis coupling is removed, each planar
$l_{ab}\times l_{ab}$ section can be integrated out in a $2D$ fashion using
(\ref{2D scaling Ansatz}).
This approximation is equivalent to assuming that fluctuations on length scale
$l_{ab}<\xi_{ab}/\xi_{c}$ in $ab$-planes are uncorrelated in the c-axis
direction.
In addition, we keep the Imry-Ma approximation for scaling of the disorder
$\Delta\left( l_{ab}\right)$ under $2D$ block transformation
\cite{Cardy-RGbook}. Indeed, a perturbative expression for the surface tension
per unit length on the scale $l$ in $2D$ RFIM \cite{Villain-RFIM_review},
$\Sigma(l)=J\left[ 1-\sigma\kappa^{2} \ln(l)\right]$, indicates that logarithmic
corrections to $\kappa\left( l\right)$ come from the effective weakening of the
inter-block interaction $J\left( l\right)$ as domain boundaries adjust to the
random field. We thus arrive at the final form of our approximate scaling
equations for block $l_{ab}\times l_{ab}\times1$,
\begin{align}
J_{c}\left(  l_{ab}\right)   &  =J_{c}l_{ab}^{2}\label{RG2 Jc}\\
J_{ab}\left(  l_{ab}\right)   &  =J_{ab}l_{ab}\sqrt{1-\sigma\kappa^{2}%
\ln\left(  l_{ab}\right)  }\label{RG2 Jab}\\
\Delta\left(  l_{ab}\right)   &  =\Delta l_{ab}.
\label{RG2 Delta}%
\end{align}
For $l_{ab}=\xi_{ab}/\xi_{c}$, the correlation lengths in the transformed model
are isotropic $\xi_{ab}\left( l_{ab}=\xi_{ab}/\xi _{c}\right) =\xi_{c}$. Since
the disorder remains isotropic, it must be that the rescaled interactions are
also isotropic. Thus, by imposing a self-consistency condition, $J_{ab}\left(
l_{ab}=\xi_{ab}/\xi _{c}\right) = J_{c}\left( \xi_{ab}/\xi_{c} \right)$, we
obtain a mapping of the bare \emph{anisotropic} RFIM (\ref{RFIM}) onto
\emph{isotropic} $3D$ RFIM with effective interactions
\vspace{-0.15in}\begin{equation}\vspace{-0.15in}
J_{ab}\left( l_{ab}=\xi_{ab}/\xi_{c}\right) =J_{c}\left( \xi_{ab}/\xi
_{c}\right)  = \left( \frac{\xi_{ab}}{\xi_{c}}\right)^{2}J_{c} \;,
\label{solvable model}%
\end{equation}
and $\Delta\left( \xi_{ab}/\xi_{c}\right) = \left( \frac{\xi_{ab}}{\xi_{c}}
\right) \Delta$. Using (\ref{RG2 Jab}),(\ref{RG2 Delta}),(\ref{solvable model})
and expression for the correlation length of the $3D$ RFIM,
$\xi_{3D}=\Lambda\left( \kappa_{c}^{-2}-\kappa^{-2}\right)  ^{-\nu}=\xi_{c}$, we
obtain
\begin{align}
\left(  \frac{J_{ab}}{\Delta}\right)  ^{2} &  =\kappa_{c}^{-2}-\left(
\frac{\Lambda}{\xi_{c}}\right)  ^{1/\nu}+\sigma\ln\left(  \frac{\xi_{ab}}%
{\xi_{c}}\right)  \label{Jab/Delta}\\
\left(  \frac{J_{c}}{\Delta}\right)  ^{2} &  =\left(  \frac{\xi_{c}}{\xi_{ab}%
}\right)  ^{2}\left[  \kappa_{c}^{-2}-\left(  \frac{\Lambda}{\xi_{c}}\right)
^{1/\nu}\right]  .\label{Jc/Delta}%
\end{align}
Substituting $\xi_{c}=1$, we obtain the crossover line $\xi _{c}\gtrless1$ in
the phase diagram,
\begin{equation}
\label{crossover}%
\frac{J_{c}}{J_{ab}} = \frac{\Delta}{J_{ab}}\; \sqrt{\kappa_{c}^{-2} -
\Lambda^{1/\nu}}\; e^{-\frac{1}{\sigma} \left[ \left( \frac{J_{ab}}{\Delta
}\right) ^{2} + \Lambda^{1/\nu}-\kappa_{c}^{-2} \right]} .
\end{equation}
With $\kappa_{c}\approx2.27$, $\Lambda\approx0.1$ and $\nu\approx1.37$ from
numerical simulations, \cite{Fisher01-RFIM3D-numerical}, equations
(\ref{Jab/Delta}),(\ref{Jc/Delta}) allow us to retrodict the effective model
parameters. For La$_{0.5}$Sr$_{1.5}$CoO$_{4}$ (Table 1, V) we find
$\frac{\Delta}{J}\approx0.85$, and $\frac{J_{ab}}{J_{c}}>53$, indeed satisfying
the relations $J_{c}\ll\Delta<J_{ab}$ under which our approximations are valid.
For La$_{0.5}$Sr$_{1.5}$MnO$_{4}$ (Table 1, IV)
$\frac{\Delta}{J_{ab}}\approx0.64$ and $\frac{J_{ab}}{J_{c}}\approx49$. In both
cases the anisotropy of the bare coupling parameters is remarkably high! Yet, we
think it is quite realistic for the layered systems. Because neighbor
$ab$-planes are shifted so that MO$_{6}$ octahedra in one plane fit in-between
those in the other (see Fig. 1(f)), and the inter-plane spacing is rather large,
breathing-type distortions of the octahedra accompanying the CO are weakly
coupled between the planes. On the other hand, octahedra within each plane form
a corner-sharing network, so the coupling of in-plane distortions is strong. In
the pseudocubic perovskite structure, the MO$_{6}$ octahedra share apical
oxygens along the $c$-axis as well (see Fig. 1(e)), and crystal fields are more
isotropic. Note that, for a system to have a LRO groundstate, interactions need
not be isotropic. In fact, the phase diagram instructs us that only very
strongly anisotropic systems will be disordered if the disorder potential is
weak (as exemplified by Eq. (\ref{PD1}), and the above estimates of
$J_{ab}/J_{c}$ in the layered compounds).

In conclusion, we argued on very general grounds that charge order observed in
half-doped manganites and their isostructural relatives is described by the
\emph{anisotropic} ($J_{c}\neq J_{ab}$) $3D$ random field Ising model
(\ref{RFIM}). We constructed the schematic phase diagram of this model, Fig. 2,
and supported its intuitive structure by perturbative RG equations
(\ref{RG1}),(\ref{RG2}). We also obtained the quantitative estimates of the
effective model parameters as a function of the measured correlation lengths
using a Migdal-Kadanoff block transformation scheme combined with a
phenomenological scaling equation (\ref{2D scaling Ansatz}) for a 2D RFIM.

Long-range charge order is observed in the pseudo-cubic materials at low
temperatures. In contrast, a disordered ground state with anisotropic
correlations is found in the layered compounds. This disparity is naturally
explained in our model: because of the strong anisotropy of the layered
materials these systems reside in the different regions of the phase diagram.
The anisotrpic correlation lengths measured in La$_{0.5}$Sr$_{1.5}$CoO$_{4}$ and
La$_{0.5}$Sr$_{1.5}$MnO$_{4}$ corroborate our conclusion. We note that a
disordered $3D$ RFIM has well known features which may be verified
experimentally. Furthermore, our analysis may be applied to thin films, where
even in pseudocubic compounds correlations in one direction are limited by the
film thickness. The latter, together with the model parameters, will then
determine the CO \emph{finite} correlation length \emph{within the film}.
Finally, we note that charge disorder in the form of dislocations, as in Fig.
\ref{LaSrCoO-ChargeOrder}(d), has inevitable effects on the superimposed orbital
and spin orders which have thus far not been elaborated.

\begin{acknowledgments}
We thank S. A. Kivelson, A. Tsvelik, F. Essler and J.P. Hill for fruitful
discussions, and acknowledge the financial support of DOE\#DE-AC02-98CH10886
(O.Z. and I.Z.) and DOE\#DE-FG03-00ER45798 (O.Z.).
\end{acknowledgments}


\begin{thebibliography}{9}
\bibitem {comment1}Among pseudocubic manganites such CO occures in some
Sr-doped and most of the Ca-doped systems, but not in La$_{0.5}$Sr$_{0.5}%
$MnO$_{3}$ and Pr$_{0.5}$Sr$_{0.5}$MnO$_{3}$.
\end{thebibliography}
\end{document}